%% file: main.tex
\documentclass[10pt,conference]{IEEEtran}
\usepackage{lineno}

\usepackage{cite}

\ifCLASSINFOpdf
   \usepackage[pdftex]{graphicx}
   \graphicspath{{figs/}}
   \DeclareGraphicsExtensions{.pdf,.jpeg,.png}
\else
   \usepackage[dvips]{graphicx}
   \graphicspath{{../figs/}}
   \DeclareGraphicsExtensions{.eps}
\fi
\usepackage[cmex10]{amsmath}
\usepackage{amsthm}

\usepackage{algorithmic}

\usepackage{array}

\ifCLASSOPTIONcompsoc
  \usepackage[caption=false,font=normalsize,labelfont=sf,textfont=sf]{subfig}
\else
  \usepackage[caption=false,font=footnotesize]{subfig}
\fi
\usepackage{url}

\usepackage{xspace}
\usepackage{amsmath}
\usepackage{xcolor}
\usepackage{tikz}
\usepackage{hyperref}
\usepackage{soul}

\newcommand{\hidecomment}[1]{}
\newcommand{\system}{Reveal\xspace}
\newcommand{\myparagraph}[1]{\noindent \textbf{#1}\xspace}
\newcommand{\highlight}[1]{{\color{black} {#1}}}

\definecolor{search}{HTML}{97A97C}
\definecolor{vis}{HTML}{7C97A9}
\definecolor{set}{HTML}{977CA9}
\definecolor{selected}{HTML}{FFCC00}

\newcommand{\squareoutlineplain}{%
    \begin{tikzpicture}[scale=0.25, baseline=0.2ex]
        \draw[thick] (0,0) rectangle (1,1);
    \end{tikzpicture}
}

\newcommand{\circleoutlineplain}{%
    \begin{tikzpicture}[scale=0.25, baseline=0.2ex]
        \draw[thick] (0.5,0.5) circle (0.5);
    \end{tikzpicture}
}

\newcommand{\filledcircle}[1]{%
    \begin{tikzpicture}[scale=0.3, baseline=0.3ex]
        \fill[#1,draw=none] (0.5,0.5) circle (0.5);
        \draw[black] (0.5,0.5) circle (0.5);
    \end{tikzpicture}%
}

\newcommand{\squarenum}[2]{\tikz[baseline=(X.base)] \node[draw, fill=#2, text=white, minimum size=1em, inner sep=0pt, text height=1ex, text depth=0ex, align=center] (X) {\scriptsize #1};}

\newcommand{\circlenum}[2]{\tikz[baseline=(Y.base)] \node[draw, fill=#2, text=white, circle, minimum size=1em, inner sep=0pt, text height=1ex, text depth=0ex, align=center] (Y) {\scriptsize #1};}

\newif\iffinal
\finaltrue
\newcommand{\cmtid}{110}

\iffinal
\else
\usepackage[switch]{lineno}
\fi

\begin{document}
%
\title{Reveal: A Provenance-Aware System for Multimodal Urban Data Exploration}


\iffinal

\author{%
\IEEEauthorblockN{%
Leonardo Ferreira\IEEEauthorrefmark{1}, 
João Rulff\IEEEauthorrefmark{2}, 
Maryam Hosseini\IEEEauthorrefmark{3},
Aline Paes\IEEEauthorrefmark{4},
Daniel de Oliveira\IEEEauthorrefmark{4},
Fabio Miranda\IEEEauthorrefmark{1},
Marcos Lage\IEEEauthorrefmark{4}
}
\IEEEauthorblockA{%
\IEEEauthorrefmark{1}University of Illinois Chicago, \IEEEauthorrefmark{2}Fundação Getúlio Vargas, \IEEEauthorrefmark{3}University of California, Berkeley, \IEEEauthorrefmark{4}Universidade Federal Fluminense}
\IEEEauthorblockA{%
{\fontfamily{qcr}\selectfont \footnotesize
lferr10@uic.edu, joao.rulff@fgv.br, maryamh@berkeley.edu, alinepaes@ic.uff.br}\\
{\fontfamily{qcr}\selectfont \footnotesize
danielcmo@ic.uff.br, fabiom@uic.edu, mlage@ic.uff.br}
}
}

\else
  \author{SIBGRAPI Paper ID: \cmtid \\ }
  \linenumbers
\fi

\maketitle

\begin{abstract}
Urban sensing initiatives increasingly produce multimodal datasets that combine visual and textual information to capture city characteristics and subjective factors such as safety, comfort, and accessibility.
Yet, analyzing such data at scale remains difficult, as manual inspection is unfeasible and automated methods struggle to interpret subjective cues and uncover relationships across diverse data modalities.
Moreover, existing workflows rarely make exploration steps explicit, hindering reproducibility.
To address these challenges, we introduce \system, a provenance-aware visual analytics system for exploring heterogeneous urban data.
\system enables users to build dataflows by combining text and image searches across coordinated views, chaining queries into pipelines, and applying set operations within an exploration diagram while capturing analytical provenance.
We evaluate \system through two case studies focused on accessibility and walkability.
%
\end{abstract}


\IEEEpeerreviewmaketitle

\input{01-introduction}
\input{03-related}
\input{04-requirements}

\input{05-system}

\input{06-taxonomy}

\input{07-case-studies}
\input{08-expert-feedback}

\input{09-conclusion}

\section*{Acknowledgments}
This work was supported by Fundação de Amparo à Pesquisa do Estado do Rio de Janeiro (FAPERJ), Conselho Nacional de Desenvolvimento Científico e Tecnológico (CNPq), and the U.S. National Science Foundation (NSF, Awards \#2320261, \#2330565, and \#2411223).

\bibliographystyle{IEEEtran}
\bibliography{example}

\end{document}

%% file: 01-introduction.tex
\section{Introduction}\label{sec:intro}
Advances in urban sensing have led to the deployment of sensor networks, generating large multimodal datasets that combine images, text, and geospatial metadata~\cite{dove2022sounds, piadyk2023streetaware}.
Yet, deriving insights at scale remains difficult: manual inspection is unfeasible, and automated methods struggle with subjective cues, heterogeneous labels, and context-dependent urban data.
Furthermore, existing workflows rarely make analysis steps explicit, hindering reproducibility. This lack of transparency is problematic in urban decision-making, where traceability is crucial for policy and stakeholder engagement~\cite{federica_civic_2025}.
For instance, assessing safety or accessibility requires analyzing thousands of images and documents to detect obstacles and reason about their spatial distribution~\cite{weld2019deep}.
Navigating these datasets requires bridging the \emph{semantic gap} between low-level visual features and high-level concepts, a task demanding human-in-the-loop approaches that integrate automated techniques with domain expertise.
\highlight{
To address these challenges, we introduce \system, a provenance-aware visual analytics system for iterative exploration of urban concepts across images, text, and metadata. Rather than supporting information retrieval, Reveal combines multimodal search, visual exploration, and set operations within an analytical workflow. 
Its architectural design not only allows users to create personalized concepts, but also enables workflow reuse across different contexts.
}
%
Co-designed with urban planning and machine learning researchers, \system helps mitigate semantic gaps by enabling experts to connect cross-modal representations.
The system features a dataflow canvas where users externalize their analysis by creating, combining, and refining nodes that represent urban concepts. Each node aggregates data generated by textual, visual, or spatial queries, along with set operations.
To support exploration, \system automatically tracks dataflow evolution, preserving the analytical rationale,
\highlight{
which serves as a semantic blueprint for reproducing and adapting analytical definitions across datasets and urban contexts.
}
\highlight{We evaluate \system through case studies on urban accessibility using data from Project Sidewalk~\cite{saha_project_2019} and walkability-related issues using street-level images from Chicago.}
Our contributions are:
\begin{itemize}
\item An interactive visual system, \system, to explore large multimodal datasets.
\highlight{
\item A multimodal provenance mechanism that captures the expert’s exploration rationale and supports the reuse of analytical workflows across datasets and contexts.
}

\item Case studies conducted in close collaboration with domain experts in urban planning.
\end{itemize}
The complete source code and a preprocessing tutorial are available at \href{https://github.com/urban-toolkit/reveal}{github.com/urban-toolkit/reveal}.

%% file: 03-related.tex
\section{Related Work}\label{sec:related}
\myparagraph{Visualization of embedding spaces.}
Embedding spaces enable visual analysis of high-dimensional \highlight{representations of} text~\cite{liu_visual_2018}, images~\cite{miranda_urban_2020}, and audio~\cite{rulff_urban_2022}.
Urban systems~\cite{miranda_urban_2020, rulff_urban_2022} use these embedding spaces to link streetscapes, city sounds, etc., to maps, but operate on a single modality.
\highlight{
Multimodal embeddings have also been used to map semantic relationships for reasoning~\cite{li_image-text_2023} and disaster management~\cite{fallon_vamd_2019}.
}
%
\system builds on this by constructing a joint embedding of images and text using CLIP and allows multimodal and spatial queries to extract semantically coherent, unlabeled urban concepts.

\myparagraph{Visualization and provenance.}
Provenance tracks computational workflows to support reproducibility and comparison~\cite{macko_provenance_2011}.
Environments like VisTrails track parameter and pipeline changes for scientific simulations~\cite{callahan_vistrails_2006}, while Curio~\cite{curio_moreira_2025} tracks dataflow pipelines over tabular urban data.
However, these systems capture provenance at the operator level rather than tracking evolving concept collections from multimodal search.
\system extends these ideas by modeling queries and set operations as nodes in an exploration diagram. This diagram records seed data, embedding selections, spatial filters, and set operations, making concept construction transparent and replayable.

\myparagraph{Visual exploration of multimodal collections.}
Several systems support interactive exploration of large multimodal collections through galleries, summaries, and linked views~\cite{miranda_urban_2020, kuchroo_multimodal_2021, rulff_urban_2022}.
For instance, ARIES enables visual organization of art collections~\cite{crissaff_aries_2018}, while other tools target acoustic corpora~\cite{rulff_urban_2022} or multi-sensor datasets~\cite{kuchroo_multimodal_2021}.
These tools typically lack tight coupling between multimodal retrieval and explicit workflow provenance or geospatial reasoning.
In contrast, \system unifies images and captions in a shared embedding space, provides multimodal and spatial queries as first-class operations, and links them via set operations to preserve analytical reasoning steps.

%% file: 04-requirements.tex
\section{Requirements \& Challenges}\label{sec:requirements}
\system emerged from long-term collaborations with urban experts~\cite{miranda_urban_2020, moreira_urban_2024, moreira_urbanite_2025, curio_moreira_2025}. 
Planners repeatedly highlighted the need to incrementally explore complex concepts in large, multimodal datasets that combine street-level images, textual annotations, and geolocation metadata.
Existing tools only partially address these needs. Urban-scale tools such as StreetVizor~\cite{shen2017streetvizor} and Urban Mosaic~\cite{miranda_urban_2020} enable exploration of large street-level image repositories but lack tightly coupled multimodal retrieval and provenance tracking.
Conversely, workflow tools like VisTrails~\cite{callahan_vistrails_2006}, VisFlow~\cite{yu_visflow_2017}, and Curio~\cite{curio_moreira_2025} excel at visual analytics pipelines but lack native support for multimodal urban data at scale.
Based on these gaps, we established five design requirements:


\myparagraph{[R1] Enable iterative multimodal concept exploration.}\highlight{The system must support text, image, and spatial queries through coordinated views, allowing users to inspect and merge results, perform  visual and spatial search refinements, and apply set operations (union, intersection, difference) over collections to create, compare, or evolve urban concepts.
}

\myparagraph{[R2] Provide fast query responses.} To maintain analytical flow, the system must deliver interactive response times and rapid visual updates during query refinement~\cite{liu_effects_2014}.

\myparagraph{[R3] Record the history of the concepts.} The system should visually record exploration activities, including queries, results, and operations, enabling users to revisit intermediate states and branch alternative explorations.

\myparagraph{[R4] Save concept sub-collections.} Users must be able to persistently save, name, and organize curated sub-collections of images, text, and metadata (e.g., named ``buckets") to share findings and resume analysis across sessions.

\myparagraph{[R5] Support geospatial metadata.} The system should provide interactive visualizations to analyze spatial patterns, helping users investigate how multimodal concepts vary across different regions.

\myparagraph{Challenges.}
Developing this system raises major challenges. Manual exploration of large multimodal datasets yields narrow views, while automated methods (e.g., embeddings or similarity search) struggle with abstract, context-dependent concepts (e.g., ``routes that feel safe").
Consequently, human expertise is essential for interpreting context and resolving semantic ambiguities. \system must therefore combine scalable indexing with rich, provenance-aware interactions to systematically capture how experts derive their findings.

%% file: 05-system.tex
\section{The \system system}\label{sec:system}
\system is a visual analytics system designed for large datasets combining images, text, and geospatial metadata. It relies on multimodal embeddings for cross-modal retrieval and dataflow provenance tracking for transparent, reproducible exploration. The system features a backend storage and query engine coupled with an interactive visual interface.
%
%

\myparagraph{Embeddings and Provenance.}\label{sec:background}
To reason across complementary signals—such as images, text, and geolocation, \system projects data into a shared semantic space of CLIP~\cite{radford_learning_2021} features. This embedding space enables text- and image-based search, interpretable projections, and zero-shot querying of unlabeled urban concepts without additional model training.
Also, \system records system states (active filters, selected subsets) and interactions (queries, spatial selections, set operations)~\cite{fekete_exploring_2020}. This provenance information is tied directly to a dataflow diagram, preserving the user's analytical reasoning and allowing them to revisit and refine their analysis.

\myparagraph{Data preprocessing.}
\system's preprocessing pipeline includes three phases. It resizes images, indexes them, and caches them as thumbnails. Also, text annotations are lemmatized, stripping stop words to extract core tokens (nouns, verbs, adjectives, adverbs). Finally, it maps bidirectional relationships between each image and its caption tokens. \textit{Embeddings generation} uses CLIP to generate three feature vectors -- image embeddings, text embeddings from extracted tokens, and joint embeddings combining full captions with their respective images. \textit{Dimensionality reduction} uses UMAP to project high-dimensional embeddings into 2D spaces, as it produced a more interpretable cluster structure than t-SNE or PCA.


\myparagraph{Backend server.}
The server features a Storage and a Query engine to support interactive query response times.
The \textit{Storage} operates in a hybrid mode. While images and metadata are hosted remotely, token relationships, 2D projections, and precomputed CLIP feature vectors are stored locally to allow rapid query execution and persistent retrieval of historical analytical states~(\textbf{R2, R3, R4}).
The \textit{Query engine} processes multimodal inputs by utilizing CLIP to compute feature vectors for real-time text, image, or joint queries~(\textbf{R1}). It calculates the cosine similarity against precomputed vectors, accelerating search efficiency~\cite{miranda_urban_2020}~(\textbf{R2}). A metadata module handles client requests to create, update, or delete saved states~(\textbf{R4}), while a geospatial module provides point-in-polygon filtering for geographical data constraints~(\textbf{R5}).

\begin{figure*}[t]
\centering
\includegraphics[width=0.95\linewidth]{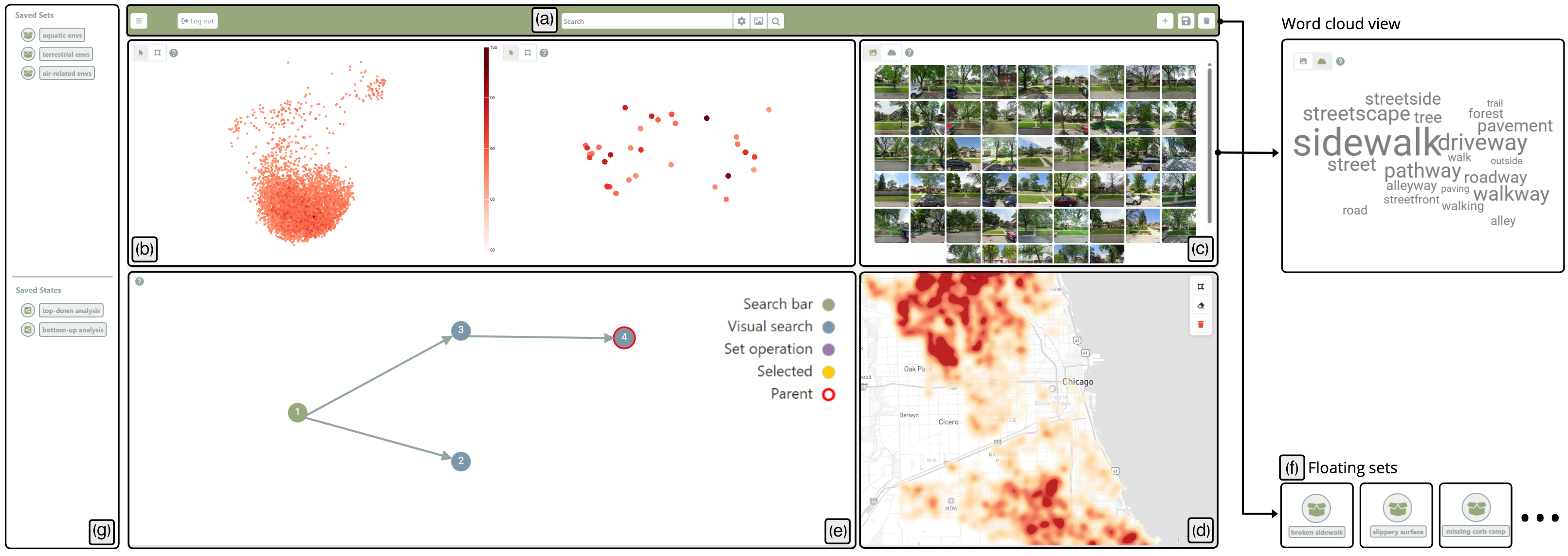}
\caption{\system interface: (a)~search bar; (b)~image and text embedding scatter plots; (c)~image gallery and token word cloud; (d)~spatial heatmap map view; (e)~dataflow exploration diagram tracking provenance; (f)~movable floating sets; (g)~saved state and collection sidebars.}
\vspace{-0.5cm}
\label{fig:interface}
\end{figure*}

\myparagraph{Visual interface.}
The interface integrates seven visual components (Figure \ref{fig:interface}) to enable multi-perspective dataset exploration.
The \textit{Search bar} \textbf{(a)} supports text inputs, target images, or joint combinations sent directly to the backend engine~(\textbf{R1}).

The \textit{Embeddings} area \textbf{(b)} displays independent 2D scatter plots for text and images or as a unified cross-modal space, exposing the global distribution and proximity of data points. The \textit{Image gallery \& word cloud} displays \textbf{(c)} retrieved image results ranked by cosine similarity. In the same view space, users can toggle a word cloud displaying the top-20 text tokens scored by similarity. The \textit{Map view} \textbf{(d)} renders an interactive heatmap showing the spatial \highlight{distribution of results~(\textbf{R5})}. The map is dynamically linked to the other views.

The \textit{Exploration diagram} \textbf{(e)} serves as the primary mechanism for multimodal provenance tracking. Each query or data transformation creates a dedicated graph node: green nodes \filledcircle{search} indicate text/image search inputs; blue nodes \filledcircle{vis} \highlight{mean visual search refinements}; and purple nodes \filledcircle{set} denote set actions (union, intersection, difference). 
\highlight{Nodes are interactive, colored yellow when selected, and numbered to strictly preserve operation order for asymmetric actions like set difference. 
}
%
Red borders designate the active parent node \highlight{(the node whose records serve as input to the next refinement)}. Users can save a complete graph path into a persistent analysis state~(\textbf{R3}).

The \textit{Floating sets} \textbf{(f)} are movable, on-screen buckets where users can drag and drop items from the gallery. They temporarily or permanently store images, texts, and metadata. \highlight{Unlike diagram nodes, floating sets are not produced by a query or a set operation. They are manually curated containers, not bound to a parent node, and a single set may aggregate items originating from different branches of the exploration.}(\textbf{R4}). The \textit{Sets and state} \textbf{(g)} sidebar stores saved data buckets and exploration graph snapshots, capturing filters, embedding thresholds, and node pipelines so users can resume analysis across sessions~(\textbf{R3, R4}).

\myparagraph{Design choices.}
\system adheres to Shneiderman’s overview-first, zoom-and-filter, then details-on-demand visual framework~\cite{shneiderman2003eyes}. The 2D embedding space serves as an abstract overview for identifying clustering trends. The gallery, word cloud, and map views function as specific context filters, enabling users to isolate micro-details or inspect spatial variance~(\textbf{R5}) while feeding elements back into the search loop~(\textbf{R1}). Conversely, the dataflow diagram abstracts the temporal dimension, tracking the workflow topology and user logic over time to guarantee reproducibility~(\textbf{R3}).

\begin{figure}[h]
\centering
\includegraphics[width=0.9\linewidth]{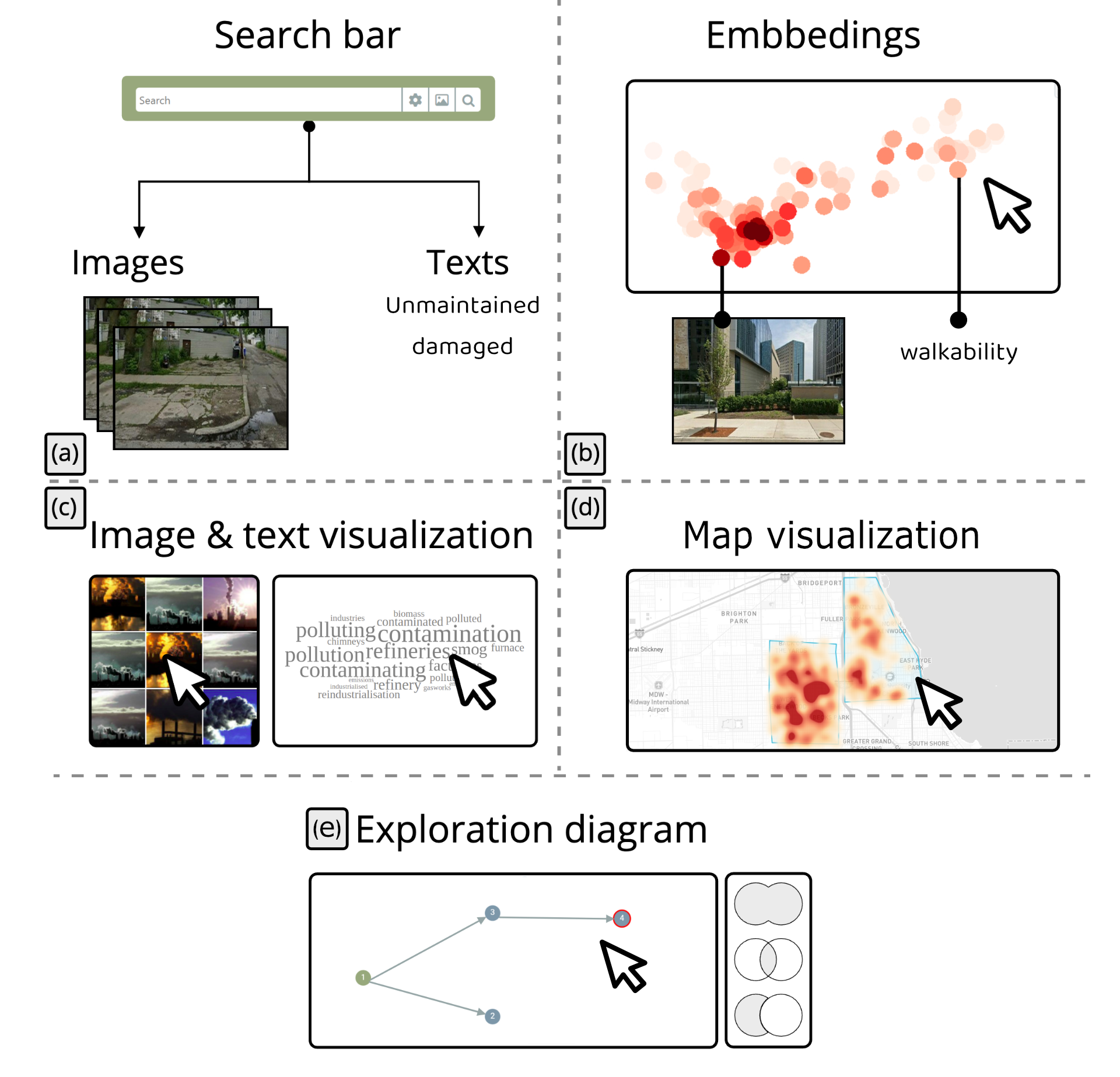}
\caption{\system offers five search interactions: (a)~multimodal search bar queries; (b)~embedding plot selections; (c)~gallery and word cloud filtering; (d)~spatial polygon filtering; and (e)~set operations of results. \highlight{All of these operations are tracked by \system's provenance system.}}
\label{fig:interactions}
\vspace{-0.25cm}
\end{figure}

%% file: 06-taxonomy.tex
\begin{figure*}[t]
\centering
\includegraphics[width=0.95\linewidth]{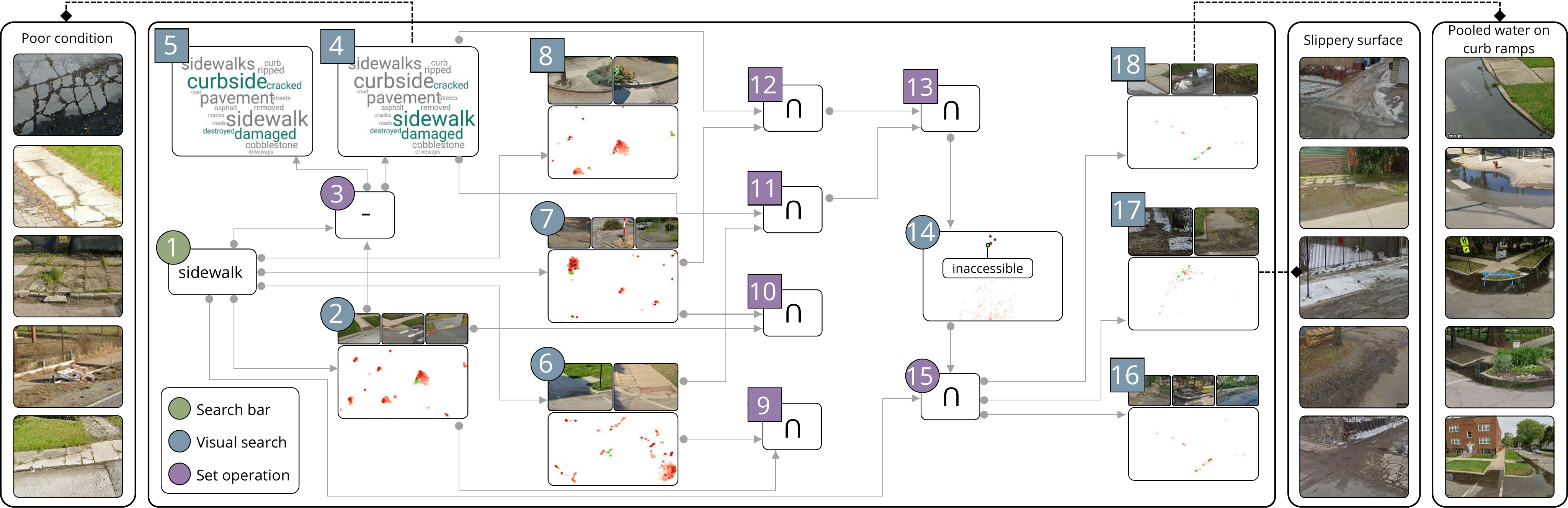}
\caption{Using \system to extend a dataset taxonomy. \protect\circleoutlineplain nodes represent baseline search steps; \protect\squareoutlineplain nodes indicate taxonomy expansions. Nodes are colored by type: \protect\filledcircle{search} (search inputs), \protect\filledcircle{vis} (canvas selections), and \protect\filledcircle{set} (set operations). This process generated 11 new granular categories missing from the original metadata. \highlight{Each node represents a set of records that can be explored through its associated interface elements (word cloud, embeddings, image gallery, etc.).}}
\vspace{-0.5cm}
\label{fig:case_study_1}
\end{figure*}

\myparagraph{Supported interactions.}\label{sec:interactions}
The interactions within \system's interface (see Figure \ref{fig:interactions}) enable users to explore a multimodal collection and extract their own datasets. To begin this exploration, \textit{Seed queries} allow users to create new nodes in the exploration diagram by entering images and text into the search bar, with the interface updating in real time as users adjust cosine similarity and date range filters. 
Once data is loaded, users can select elements from \textit{scatter plots} to analyze individual or multiple data points, which can initiate new searches~(\textbf{R1}), visualize cross-modal links by highlighting corresponding text-image pairs, and update the matching items within the image gallery and word cloud. 
Similarly, analysts can select elements from \textit{image galleries and word clouds} to trigger new queries, highlight their matching vector locations in the scatter plots, or drag specific images into sets to automatically save them along with their associated textual metadata~(\textbf{R4}). To incorporate geographical constraints, users can select elements based on their \textit{location} by drawing custom polygons directly onto the Map View~(\textbf{R5}), an action that automatically generates a new node in the exploration diagram containing only the geographically bounded elements from the parent node to enable localized analysis and regional comparisons. Finally, the system allows users to combine and recover \textit{exploration states} via the interactive diagram, where clicking nodes reveals their underlying query parameters, new search bar inputs generate fresh analytical branches, canvas interactions link parent nodes to children, and set operations connect child nodes to both parents while visually tagging the interface components used to guarantee full workflow traceability (Figures \ref{fig:interface}(e) and \ref{fig:interactions}(e)).

%% file: 07-case-studies.tex
\begin{figure*}[!t]
\centering
\includegraphics[width=0.95\linewidth]{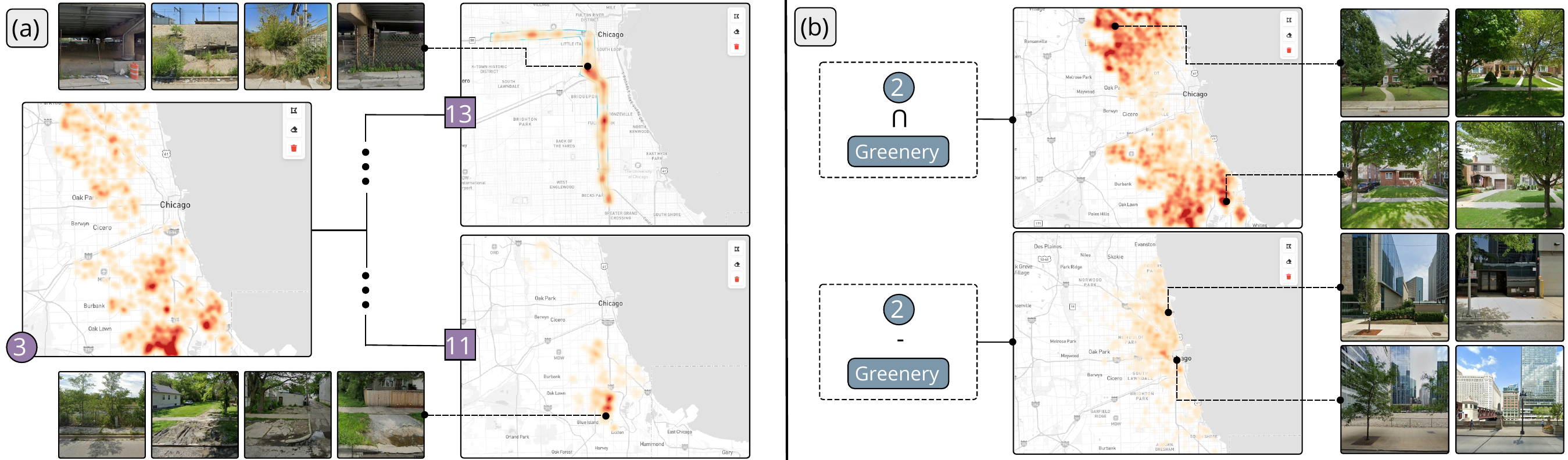}
\caption{Spatial analysis of walkability and accessibility in Chicago using an LLM-enriched street-level dataset. (a) Mapping patterns of structural decay from generic issues to highly localized infrastructure corridors and active construction zones. (b) Isolating high-quality urban environments, distinguishing between downtown hardscapes and residential green walkability.}
\vspace{-0.5cm}
\label{fig:new_case_study}
\end{figure*}

\section{Case Studies}\label{sec:case-studies}
This section presents two case studies evaluated by an urban planner with over 11 years of experience. First (Section~\ref{sec:taxonomy}) the expert used \system to explore the Project Sidewalk dataset~\cite{saha_project_2019} and expand its taxonomy by identifying unlabeled accessibility issues. 
Second (Section~\ref{sec:walkability}), the expert applied this expanded taxonomy to a new dataset of geolocated Chicago street-level images, thereby reconstructing an analogous analysis of structural decay and walkability.
%
Nodes are referred to as follows: \circlenum{}{search} for search bar queries, \circlenum{}{vis} for visualization selections, and \circlenum{}{set} for set operations.
\subsection{Expanding Project Sidewalk's taxonomy}\label{sec:taxonomy}
Project Sidewalk~\cite{saha_project_2019} crowdsources sidewalk accessibility information via Google Street View images, providing labels for missing curb ramps, obstacles, and surface problems. While broad, this taxonomy misses nuanced urban conditions. 
\system leverages multimodal search combining image and text queries (\textbf{R1}) to detect overlooked issues and formalize them into a detailed taxonomy. Nodes shown as \circleoutlineplain keep the original taxonomy, whereas \squareoutlineplain indicate taxonomy expansions.

The study began with a keyword query for ``sidewalk'' (Figure~\ref{fig:case_study_1}, \circlenum{1}{search}). The expert identified a cluster of well-preserved sidewalks and selected three representative points to create \circlenum{2}{vis}. A set difference between \circlenum{1}{search} and \circlenum{2}{vis} generated \circlenum{3}{set}, isolating poorly maintained sidewalks. \highlight{From the word cloud of node \circlenum{3}{set}, the expert selected descriptors to construct \squarenum{4}{vis} (cracked, damaged, destroyed, and sidewalks) and \squarenum{5}{vis} (cracked, damaged, destroyed, and curbside).}
Returning to \circlenum{1}{search}, the expert isolated three distinct curb ramp configurations from the embedding scatter plot: sidewalks with tactile pavement (\circlenum{6}{vis}), sidewalks with a curb ramp but no tactile pavement (\circlenum{7}{vis}), and sidewalks lacking both features (\squarenum{8}{vis}). Re-visiting the well-maintained sidewalks (\circlenum{2}{vis}), the expert intersected it with \circlenum{6}{vis} and \circlenum{7}{vis} to generate \squarenum{9}{set} (well-maintained with tactile pavement) and \squarenum{10}{set} (well-maintained without tactile pavement). Similarly, intersecting the damaged sidewalks (\squarenum{4}{vis}) with \circlenum{6}{vis} and \circlenum{7}{vis} yielded \squarenum{11}{set} (tactile pavement in poor condition) and \squarenum{12}{set} (curb ramps in poor condition). Intersecting these two nodes created \squarenum{13}{set} (both tactile pavement and curb ramps in poor condition).
Further embedding exploration of \squarenum{13}{set} surfaced the term ``inaccessible,'' which was used to create \circlenum{14}{vis} and subsequently intersected with \circlenum{1}{search} to create \circlenum{15}{set}. 
\highlight{From the embeddings scatter plots of images of \circlenum{15}{set}, the expert extracted three final clusters: sidewalks under construction (\squarenum{16}{vis}), slippery surfaces (\squarenum{17}{vis}), and water accumulation on curb ramps (\squarenum{18}{vis})}.
In total, 11 detailed categories (\squarenum{4}{vis}, \squarenum{5}{vis}, \squarenum{8}{vis}, \squarenum{9}{set}, \squarenum{10}{set}, \squarenum{11}{set}, \squarenum{12}{set}, \squarenum{13}{set}, \squarenum{16}{vis}, \squarenum{17}{vis}, \squarenum{18}{vis}) were established, aggregating over 1,600 newly labeled images (\textbf{R3, R4}). These granular conditions offer a more detailed breakdown than Project Sidewalk's broader tags (e.g., missing tactile ``warning'' or ``uneven pavement'').

\subsection{Urban Accessibility and Walkability in Chicago}\label{sec:walkability}
The expert transposed this expanded taxonomy to a dataset of geolocated Chicago Google Street View images enriched with LLM-generated captions. Using the previous dataflow graph as a semantic blueprint, the expert replicated the taxonomy's filtering logic to explore structural decay and walkability patterns across Chicago.

\myparagraph{Investigating structural decay.} The expert reproduced the Chicago counterpart to \circlenum{3}{set} (poorly maintained sidewalks). As a generic category, its spatial projection spanned most neighborhoods, with a slightly higher density in South Chicago and minimal instances in the Loop (Figure~\ref{fig:new_case_study}(a), \circlenum{3}{set}). To isolate discriminative spatial trends, the expert replicated the refinement for \squarenum{11}{set} \highlight{(i.e., from node \circlenum{1}{search} produced node \circlenum{6}{vis} and from node \circlenum{3}{set} generated node \circlenum{4}{vis}, whose intersection is node \squarenum{11}{set}, which contains records of pavements in poor condition)}, which revealed a distinct cluster in South Chicago, characterized by eroded curb edges and mud-covered segments (Figure~\ref{fig:new_case_study}(a), \squarenum{11}{set}).
\highlight{Similarly, reconstructing the filtering steps used to select the records that compose node \squarenum{13}{set} (tactile pavement and curb ramps in poor condition) via a node intersection exposed a linear spatial pattern tightly bound to major transportation corridors (e.g., highways, rail lines seen in Figure~\ref{fig:new_case_study}(a), \squarenum{13}{set}).} Area-based queries using the map confirmed that these infrastructure corridors consistently suffered from compounded degradation. Finally, replicating the ``under construction'' node (\squarenum{16}{set}) isolated active municipal maintenance zones, successfully separating temporary disruptions from areas of persistent structural decay.

\myparagraph{Investigating walkable urban environments.} The expert reconstructed the well-maintained branch, beginning with the Chicago analogue to \circlenum{2}{vis}. This produced a dense concentration of points in downtown Chicago, corresponding to the low incident rate of structural issues in the area.
To evaluate environmental attributes, the expert intersected this well-maintained node with text tokens representing trees and street greenery, applying a branching strategy analogous to the one used for \squarenum{9}{set} and \squarenum{10}{set}. The resulting spatial heatmap inverted the downtown pattern: the Loop largely disappeared, and two prominent clusters emerged in northern and southern residential zones (Figure~\ref{fig:new_case_study}(b)). Image inspection confirmed that these clusters captured neighborhoods combining well-maintained pathways with significant street greenery, effectively distinguishing a residential ``green walkability'' typology from downtown Chicago's standard urban hardscape.

%% file: 08-expert-feedback.tex
\section{Experts' Feedback}\label{sec:feedback}

To evaluate the system's capabilities, we conducted a semi-structured interview with the urban expert after completing both case studies.
The discussion focused on three themes: \textit{the value of multimodal querying}, \textit{the role of provenance in structuring analytical workflows}, and \textit{the scalability of the approach across different urban contexts}.

\myparagraph{Bridging the semantic gap.}
The expert highlighted that combining image and text-based search improved the identification of nuanced accessibility issues.
While text-based queries were effective for retrieving broad categories, image-based queries enabled the discovery of ``visually similar patterns that are difficult to describe with keywords alone,'' such as the characteristic appearance of muddy or abandoned sidewalks in South Chicago.
The expert also noted that set operations were key for refining these visual clusters into distinct semantic concepts.
%
For instance, the expert used intersection operations to combine degraded tactile pavement clusters and curb ramps, then applied difference operations with the construction-related node to separate long-term decay from active work zones.

\myparagraph{Provenance as an analytical blueprint.}
The expert identified the exploration graph as a feature of high importance.
Unlike tools in which the analytic history is obscured or lost, \system's provenance diagram provided a ``visual audit trail'' that made the rationale behind each derived category explicit.
According to the expert, this transformed the workflow from a linear sequence of searches into an iterative process in which branches could be tested, adjusted, or pruned without losing the broader context.
Such transparency is particularly important, since justifying \textit{why} a specific region is flagged as ``inaccessible'' is as important as the classification itself.

\myparagraph{Reproducibility and transposition.}
The expert underscored the value of using the provenance structure from the first case study as a ``semantic blueprint'' for the Chicago analysis.
Typically, applying a taxonomy to a new city requires restarting the exploration due to differences in data distribution and visual characteristics.
By reconstructing the same semantic operations, the expert ensured that the definition of ``walkability'' applied to Chicago was methodologically aligned with the criteria established in the Project Sidewalk analysis.
According to the expert, this functionality can be used in comparative urban analysis, as it provides a methodological basis for applying nuanced and subjective definitions of urban ``quality'' across multiple cities.

%% file: 09-conclusion.tex
\section{Conclusion and Future Work}\label{sec:conclusion}
In this work, we have presented \system, a provenance-aware system that enables the exploration of multimodal data composed of images and text.
\system enables users to externalize their data exploration process through a dataflow diagram. This diagram, together with a set of visualizations, were specifically designed to integrate image and text data \highlight{
and serve as a semantic blueprint for reproducing and adapting analytical definitions across contexts.
}
Building on previous works~\cite{miranda_urban_2020, rulff_urban_2022}, \system provides an approach for both interactive and iterative exploration of a dataset, allowing users to operate on sets for data analysis.
In future work, we plan to explore additional multimodal representation learning methods beyond CLIP to further enhance the conceptual understanding of images and text. Emerging models, such as BLIP \cite{li_blip_2022}, have exhibited promising results, offering potential avenues for improvement in the analysis and comprehension of multimodal data. 
We also plan to explore the use of \system as an educational tool to help instructors break down concepts for students.
This is aligned with previous work~\cite{silva_using_2010} that leveraged dataflow-based visualizations to introduce students to scientific concepts.
Finally, together with data from our other systems that also have provenance capabilities~\cite{de2023p,curio_moreira_2025}, we plan to conduct large-scale studies to better understand the analysis paths users follow during their exploratory data analysis process.
%